\documentclass{aa}

\usepackage{txfonts}
\usepackage{natbib,twoopt}
\usepackage{graphicx}
\usepackage{arydshln}
\usepackage{enumitem}
\usepackage{multirow}
\usepackage{multicol}
\usepackage{longtable}
\usepackage{hyperref}
\usepackage{url}
\usepackage{color}
\bibpunct{(}{)}{;}{a}{}{,} 
\usepackage{arydshln}
\usepackage{lastpage}

\hypersetup{
	pdfauthor={A. J. Gallagher},
	pdfcreator={A. J. Gallagher},
		colorlinks, 
		citecolor=blue, 
		filecolor=blue, 
		linkcolor=blue,
		breaklinks=true, 
		plainpages=false,
		urlcolor=blue
		}
\title{Observational constraints on the origin of the elements}
\subtitle{II. 3D  non-LTE formation of \ion{Ba}{ii} lines in the solar atmosphere}

\author{
A. J. Gallagher\inst{1}
\and
M. Bergemann\inst{1}
\and
R. Collet\inst{2}
\and
B. Plez\inst{3}
\and
J. Leenaarts\inst{4}
\and
M. Carlsson\inst{5,6}
\and
S. A. Yakovleva\inst{7}
\and
A. K. Belyaev\inst{7}
}

\institute{
Max-Planck-Institut f{\"u}r Astronomie, K\"{o}nigstuhl 17, 69117 Heidelberg, Germany.
\and
Department of Physics and Astronomy, Aarhus University, Ny Munkegade 120, DK-8000 Aarhus, Denmark.
\and
LUPM, UMR 5299, Universit\'{e} de Montpellier, CNRS, 34095 Montpellier, France.
\and
Institute for Solar Physics, Department of Astronomy, Stockholm University, AlbaNova University Centre, SE-106 91 Stockholm, Sweden.
\and
Rosseland Centre for Solar Physics, University of Oslo, P.O. Box 1029 Blindern, NO-0315 Oslo, Norway.
\and
Institute of Theoretical Astrophysics, University of Oslo, P.O. Box 1029 Blindern, NO-0315 Oslo, Norway.
\and
Department of Theoretical Physics and Astronomy, Herzen University, St Petersburg 191186, Russia.
}

\date{Received ... / Accepted ...}

\authorrunning{A. J. Gallagher et al.}
\titlerunning{3D non-LTE Ba in the Sun}

\newcommand{\aba}{A({\rm Ba})}
\newcommand{\fodd}{f_{\rm odd}}
\newcommand{\kms}{{\rm km\,s^{-1}}}
\newcommand{\feh}{{\rm [Fe/H]}}
\newcommand{\teff}{T_{\rm eff}}
\newcommand{\logg}{\log{g}}
\newcommand{\stagger}{{\sc stagger}}
\newcommand{\tda}{$\langle{\rm 3D}\rangle$}
\newcommand{\multitd}{MULTI3D}
\newcommand{\multi}{MULTI 2.3}
\newcommand{\logtr}{\log{\tau_{\rm ROSS}}}
\newcommand{\vsini}{v\sin{i}}
\newcommand{\hi}{\ion{H}{i}}

\newcommand{\fei}{\ion{Fe}{ii}}

\newcommand{\bai}{\ion{Ba}{i}}
\newcommand{\baii}{\ion{Ba}{ii}}
\newcommand{\baiii}{\ion{Ba}{iii}}
\newcommand{\opd}{\log{\tau_{\rm 5000}}}
\newcommand{\sigph}{\sigma_{\rm photo}}

\definecolor{grey}{rgb}{0.5,0.5,0.5}
\definecolor{orange}{rgb}{1.0,0.5,0.0}
\definecolor{brightblue}{rgb}{0.7,0.9,1.0}
\definecolor{brightgreen}{rgb}{0.8,1.0,0.8}
\definecolor{brightblue}{rgb}{0.7,0.9,1.0}
\definecolor{brightgreen}{rgb}{0.8,1.0,0.8}
\definecolor{violet}{rgb}{1.0,0.0,1.0}

\newcommand{\Ba}[5]{\mbox{$#1\,^#2{\rm #3}^{{\rm #4}}_{\rm #5}$}}

\defcitealias{Bergemann2019}{Paper I}
\newcommand{\po}{\citetalias{Bergemann2019}}

\abstract
{The pursuit of more realistic spectroscopic modelling and consistent abundances has led us to begin a new series of papers designed to improve current solar and stellar abundances of various atomic species. To achieve this, we have began updating the three-dimensional (3D) non-local thermodynamic equilibrium (non-LTE) radiative transfer code, \multitd, and the equivalent one-dimensional (1D) non-LTE radiative transfer code, \multi.}
{We examine our improvements to these codes by redetermining the solar barium abundance. Barium was chosen for this test as it is an important diagnostic element of the s-process in the context of galactic chemical evolution. New \baii\ + H collisional data for excitation and charge exchange reactions computed from first principles had recently become available and were included in the model atom. The atom also includes the effects of isotopic line shifts and hyperfine splitting.}
{A grid of 1D LTE barium lines were constructed with \multi\ and fit to the four \baii\ lines available to us in the optical region of the solar spectrum. Abundance corrections were then determined in 1D non-LTE, 3D LTE, and 3D non-LTE. A new 3D non-LTE solar barium abundance was computed from these corrections.}
{We present for the first time the full 3D non-LTE barium abundance of $\aba=2.27\pm0.02\pm0.01$, which was derived from four individual fully consistent barium lines. Errors here represent the systematic and random errors, respectively.}
{}


\begin{document}

\maketitle

\section{Introduction}
\label{sec:intro}

Barium is key element that is used in heavy element studies in stars. Its abundance patterns in the halo, in field stars, and in clusters have been carefully measured over the past several decades. Barium, like most other heavy elements, mostly forms via a series of neutron captures through either the rapid (r-) process or slow (s-) process channels. These two neutron capture channels have very different sites. After the discovery and analysis of 2017gfo \citep[the electromagnetic counterpart of GW170817][]{Valenti2017}, it is highly probable that the r-process mostly occurs in neutron star mergers \citep{Thielemann2011}. Conversely, the majority of barium in the Sun \citep[$81\%$][]{Arlandini1999} ostensibly formed via the s-process in thermally-pulsing asymptotic giant branch (TP-AGB) stars \citep{Smith1988}. However, other sites for the s-process and r-process do and most likely exist. Naturally, the barium isotope ratio, $\fodd$\footnote{$\fodd\equiv\left[ N(\element[][135]{Ba}) + N(\element[][137]{Ba})\right]/N(\element{Ba})$}, of a star is a useful quantity as it provides precise information on the s- and r-process contribution, but is exceedingly difficult to measure. Therefore, this information is only measured in some thick disk and halo stars, where this parameter is most interesting \citep{Magain1993,Magain1995,Mashonkina1999,Gallagher2010,Gallagher2012,Gallagher2015}.

Most abundances, save those such as lithium that are measured in absolute units, are measured relative to the solar abundances. This helps to mitigate systematic errors within spectroscopic abundances and yields extra information about stellar populations, evolutionary stages and ages, that measurements in absolute units might not. As a result, the solar abundances are extremely important to stellar astrophysics. Consequently, very accurate measurements of the solar abundance are needed, which employ sophisticated model atmospheres and spectrum synthesis techniques such as 3D hydrodynamics and non-local thermodynamic equilibrium (non-LTE) physics. In recent years, with the development of faster and larger computers, it has been possible to develop and implement these methods \citep{Asplund2003,Steffen2015,Klevas2016,Amarsi2016o,Mott2017,Amarsi2017si,Nordlander2017}. 

One of the main aims of this paper series is to report on our development of the one dimensional (1D) and three-dimensional (3D) statistical equilibrium codes -- \multi\ \citep{Carlsson1986} and \multitd\ \citep{Leenaarts2009} -- as we include new or better physics into their program flows. Given how important barium is to galactic chemical evolution studies because it traces the impact of neutron-capture nucleosynthesis, we present a thorough analysis of the solar barium abundance using a handful of \ion{Ba}{ii} optical lines computed using the two statistical equilibrium codes and the same barium model atom. 

The statistical equilibrium of \baii\ has already been a subject of several detailed studies \citep{Mashonkina1996, Mashonkina1999, Shchukina2009, Andrievsky2009, Korotin2015}. The first such study was conducted by \citet{Gigas1988} in Vega. There are, however, important differences between our work and these earlier studies. First, we use the new quantum-mechanical rates for transitions caused by inelastic collisions with hydrogen atoms from \citet{Belyaev2018}. We also examine the impact dynamical gas flows have on \baii\ by utilising a 3D radiative hydrodynamical model to compute full 3D non-LTE radiative transfer, as well as 3D LTE, 1D LTE, and 1D non-LTE. Ab initio collisional damping from \citet{Barklem2000} was included in the linelist. 

It has been observationally confirmed that the \baii\ resonance line at 4554\,\AA\ is sensitive to the chromospheric effects\footnote{both FAL-C semi-empirical models and a 3D radiative hydrodynamical model from \citet{Asplund2000} were used.}, and so naturally a polarised spectrum of the resonance line is also sensitive to the quantum interferences \citep[see, e.g.][]{Kostik2009,Shchukina2009,Belluzzi2013,Smitha2013,Kobanov2016}, however, this is beyond the scope of this paper.

The paper is structured as follows. In Sect.~\ref{sec:model} we describe the observations, we detail the model atmospheres, model atoms and spectral synthesis codes; in Sect.~\ref{sec:lineformation} we discuss the impact that various model assumptions have on our results; in Sect.~\ref{sec:results} we describe the analysis and results from our \baii\ line analysis; and in Sect.~\ref{sec:conclusions} we summarise the study.

\section{Models and Observations}
\label{sec:model}

\subsection{Solar spectrum}
\label{sec:observations}

The solar spectrum is taken from the Kitt Peak National Observatory (KPNO) solar atlas published by \citet{Kurucz1984}. This solar atlas covers the spectral range of $3\,000$ to $13\,000$\,\AA\ at a typical resolution ${\rm R}\equiv\frac{\lambda}{\Delta\lambda}=400\,000$. Although newer solar spectra exist such as the PEPSI spectrum provided by \citet{Strassmeier2018}, we chose to work with the former atlas as it has a very high resolution, roughly twice that of the latter. Nevertheless, comparisons of these two spectra have previously been made and they were found to be in very good agreement with one-another \citep{Osorio2019}.

\subsection{1D model atmosphere}
\label{sec:1Datmos}

We use the MARCS model atmosphere that was computed for the Sun from the opacity sampled grid published in \citet{Gustafsson2008}. The solar parameters of this model are $\teff/\logg/\feh=5777/4.44/0.00$ and include a mixing length parameter, $\alpha_{\rm MLT}=1.50$. The solar composition used to compute the model opacities are based on those published in \citet{Grevesse2007}.

\subsection{3D model atmosphere}
\label{sec:3Datmos}

For the work presented in this study we make use of the solar \stagger\ \citep{Nordlund1994,Nordlund1995} model with stellar parameters $\teff/\logg/\feh=5777\,{\rm K}/4.44/0.0$, from the \stagger\ model atmosphere grid \citep{Collet2011,Magic2013a}. A 3D model consists of a series of computational boxes that represent a time series, which are commonly referred to a snapshots. These snapshots are selected from a larger time series of snapshots that are produced from the \stagger\ code and are selected at a time when the simulation has reached dynamical and thermal relaxation. For our purposes -- and for the sake of time -- we have chosen to work with five snapshots, each consisting of $240\times240\times230$ grid points which cover a geometrical volume of $7.96\times7.96\times3.65$\,Mm in the $x$, $y$, and $z$ dimensions, respectively. These are used as independent input for our statistical equilibrium code \multitd\ (Sect.~\ref{sec:multi3d}) and then the output are averaged together, therefore applying the ergodic approximation that averaging in time is equivalent to averaging over space. In this case it is assumed that averaging in time is equivalent to averaging across the disc of the Sun. The emergent fluxes from these snapshots have an equivalent width variance of only $\sim0.75$\,m\AA, suggesting that including further snapshots to the study will not greatly improve the results presented in this study, only increase the computational times.

Line opacities were collected from the MARCS database and are sorted into 12 opacity bins. Continuous absorption and scattering coefficients are taken from \citet{Hayek2010}. Importantly, and unlike an equivalent 1D model, 3D models provide $x$, $y$, and $z$ velocity fields for every voxel meaning that post-processing spectrum synthesis codes provide more accurate approximations for the Doppler broadening, including asymmetric line profiles, which result from these gas flows.

We also make use of the averaged 3D model to help make qualitative comparisons between the full 3D and 1D models, however, we do not use it with \multi\ or \multitd. A \tda\ model is computed from a 3D model by spatially averaging the thermal structure of the 3D computational box over surfaces of equal Rosseland optical depth. As this can be performed in several different ways, comparing results from different studies that do not specify their averaging techniques is ultimately self-defeating.

\begin{figure}
\includegraphics[width=\linewidth]{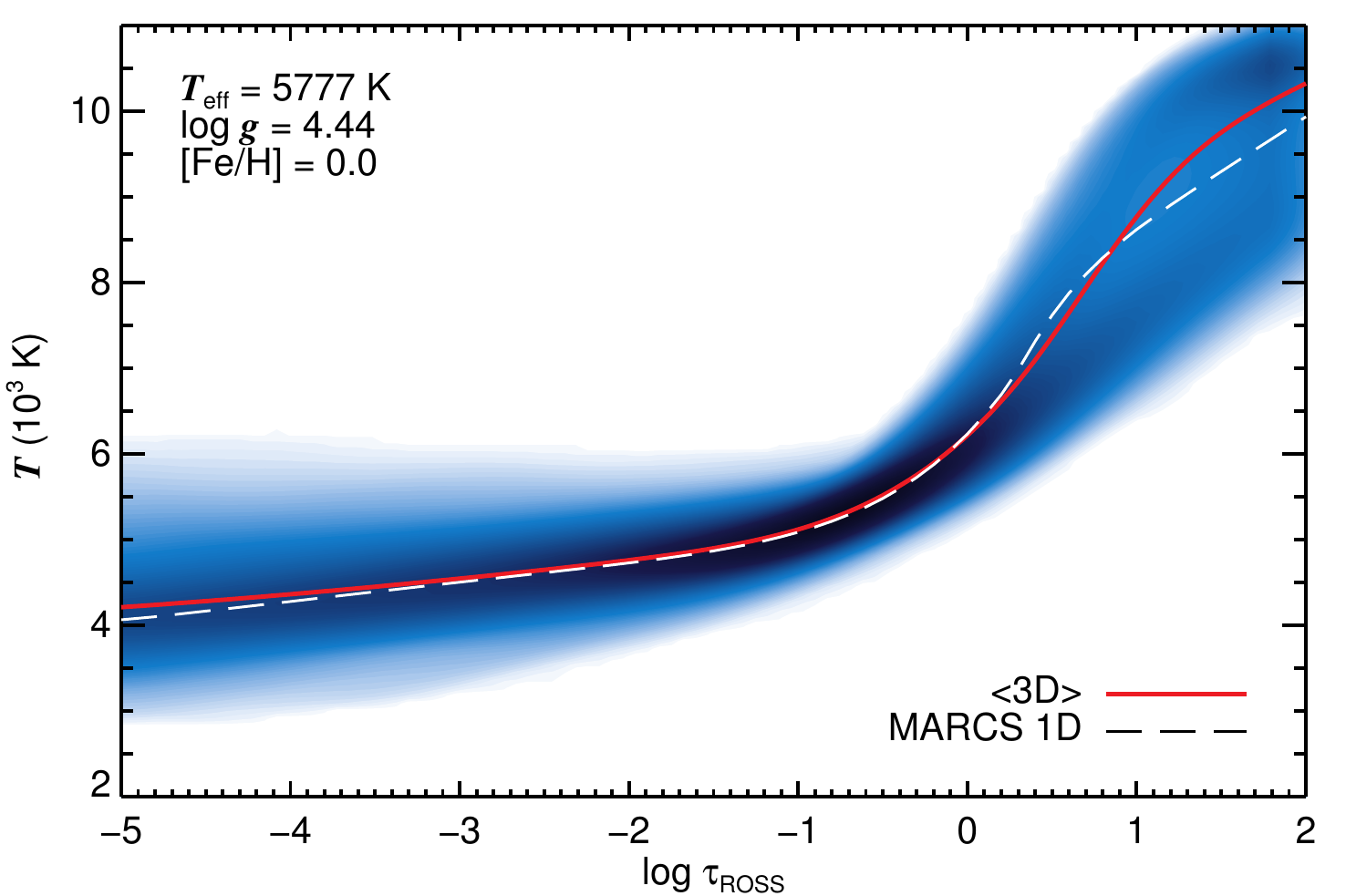}
\caption{3D,1D and \tda\ temperature structure.}
\label{fig:tstruct}
\end{figure}

Figure~\ref{fig:tstruct} depicts the 3D solar temperature structure (blue 2D histogram), along with the 1D MARCS (dashed line) and \tda\ (solid red line) temperature structures. It is clear that the average temperature of the full 3D model and the 1D model are fairly consistent in the outermost regions of the atmosphere (as seen by comparing the 1D with the \tda\ model). However, in deeper regions of the models -- where the continuum usually forms ($\logtr\approx0$) -- the models begin to diverge. This is mostly due to the differences between the convection indicative in the 3D hydrodynamic model atmosphere -- which the \tda\ model traces -- and the treatment of convection theory (in this case the mixing length theory) in the 1D model atmosphere.

\subsection{\multi}
\label{sec:multi1d}

MULTI  solves the equations of radiative transfer and statistical equilibrium in 1D geometry with 1D model atmospheres. The latest release of MULTI is \multi. However, we have made several minor changes to \multi\ for our purposes including, the ability to compute the detailed balance for charge transfer processes between ions and hydrogen. We include a fixed microturbulence value of $1\,\kms$ in our computations with the solar MARCS model. The flux data were computed using five $\mu$-angles, assuming a Gaussian quadrature scheme taken from \citet{Lowan1942}.

\subsection{\multitd}
\label{sec:multi3d}

\begin{table*}[h]
\caption{The \baii\ lines used in the abundance analysis of the Sun}
\begin{center}
\setlength{\tabcolsep}{0.01\linewidth}
\begin{tabular}{l c c c c c c c}
\hline
\noalign{\smallskip}
   Wavelength [$\AA$] & Elow (eV) & Eup (eV) & conf & conf & $\log{gf}$ & VdW  & EW (m\AA) \\ 
\noalign{\smallskip}
\hline\hline
\noalign{\smallskip}
   4554.033 & 0.00 & 2.72 & \Ba{6s}{2}{S}{}{0.5} & \Ba{6p}{2}{P}{\circ}{1.5} &  \phantom{$-$}$0.170\pm0.004$ & 303.222 & 207 \\
   5853.675 & 0.60 & 2.72 & \Ba{5d}{2}{D}{}{1.5} & \Ba{6p}{2}{P}{\circ}{1.5} & $-1.023\pm0.005$ & 365.264 & 68 \\
   6141.713 & 0.70 & 2.72 & \Ba{5d}{2}{D}{}{2.5} & \Ba{6p}{2}{P}{\circ}{1.5} & $-0.070\pm0.005$ & 365.264 & 126 \\
   6496.898 & 0.60 & 2.51 & \Ba{5d}{2}{D}{}{1.5} & \Ba{6p}{2}{P}{\circ}{0.5} & $-0.365\pm0.004$ & 365.264 &  102 \\
\noalign{\smallskip}
\hline
\end{tabular}
\end{center}
\tablefoot{The wavelengths are given in air. The equivalent widths correspond to the measurements in the solar KPNO flux atlas and are given in m\AA. The Van der Waals broadening parameters are taken from \citet{Barklem2000}. The $\log{gf}$ values reported here are derived from the very accurate transition probabilities taken from \citet{DeMunshi2015} and \citet{Dutta2016}.}
\label{tab:config}
\end{table*}

\multitd\ is an message passing interface (MPI)-parallelised, domain-decomposed 3D non-LTE radiative transfer code that solves the equations of radiative transfer using the Multi-level Accelerated Lambda Iteration (ALI) method \citep{Rybicki1991,Rybicki1992} for 3D model atmospheres. Every element that is modelled by \multitd\ is assumed to have no effect on the model atmosphere, as it is in \multi. This is a good assumption for barium as it is not an electron donor nor does it have a high impact on the overall opacity, unlike magnesium or iron, for example.

At present, it will accept three types of 3D model atmospheres formats as direct input, including those computed using Bifrost \citep{Gudiksen2011}, and \stagger. While Bifrost models are read using MPI IO, the \stagger\ models are, at present, not read this way due to complications in converting byte ordering. However, the added delay to the code's run time is minimal, and only becomes noticeable when \multitd\ is run on several hundred CPUs. In addition to these two types of model atmosphere, the code will also accept any 3D model formatted so that the temperature, $T$, density, $\rho$, electron number density, $n_{\rm e}$, and $x$, $y$, and $z$ velocity fields are supplied on a Cartesian grid that is both horizontally periodic and equidistantly spaced. Therefore, it is relatively straightforward to convert almost any 3D model to this input format for \multitd. 

We have introduced new coding for computing fluxes inside \multitd\ along with the appropriate post-processing routines designed to extract the flux data. All of the output flux data computed for the work presented here was calculated using a Lobatto quadrature scheme and the appropriate corresponding weights \citep{Abramowitz1972}. At a later stage of this paper series, other quadrature schemes will be introduced, as well as internal routines that will compute fluxes inside \multitd\ and write them as output. 

\multitd\ is now capable of accepting model atoms that include hyperfine structure (HFS) and isotope shift information for any atomic transition. This means that lines with highly asymmetric profiles, caused by these effects, can now be adequately modelled by \multitd. To test this upgrade, and to test that we could limit the impact of systematic errors dominating the abundances and abundance corrections we provide, we compared 1D spectra computed by both \multi\ and \multitd. This was conducted only for the vertical intensity ($\mu=1$), using a small test barium model atom, under the assumption of LTE. We use the same opacity sources, and same input model atmosphere. Systematic differences in the equivalent widths of less than $2.2$\,\% were found between intensities computed with \multi\ and intensities computed with \multitd. This translates to abundances differences much less than 0.01\,dex. The reason for these small differences is likely because of the way each code solves the radiative transfer equation; \multitd\ uses a direct 1D integration of the radiative transfer equation when computing spectra from 1D model atmospheres\footnote{ordinarily \multitd\ uses a short-characteristic solver for 3D model atmospheres.}, while \multi\ utilises a faster Feautrier method. However, abundance uncertainties found here are far smaller than the errors we report in Sect.~\ref{sec:results}. Therefore, we were satisfied that comparing 1D output from \multi\ with 3D output from \multitd\ was adequate.

We ran \multitd\ in short characteristic 3D solver mode and used the solar \stagger\ model as input. The \stagger\ model's $xy$ grid points were scaled down by a factor of 64 from $240\times240$ to $30\times30$ grid points using a simple bilinear interpolation scheme. Significant tests conducted in the first paper in this paper series, \citep[][henceforth, \po]{Bergemann2019}, revealed no significant loss of information in the horizontal gas flows that affected the line profiles in any noteworthy way. The horizontal components were also assumed to be periodic so that rays with very low $\mu$ angles could be computed without encountering a horizontal boundaries. The vertical grid size remained consistent with the original model atmosphere at 230 grid points.

\subsection{Model of Barium atom}
\label{sec:atom}

\begin{figure}
\includegraphics[width=\linewidth]{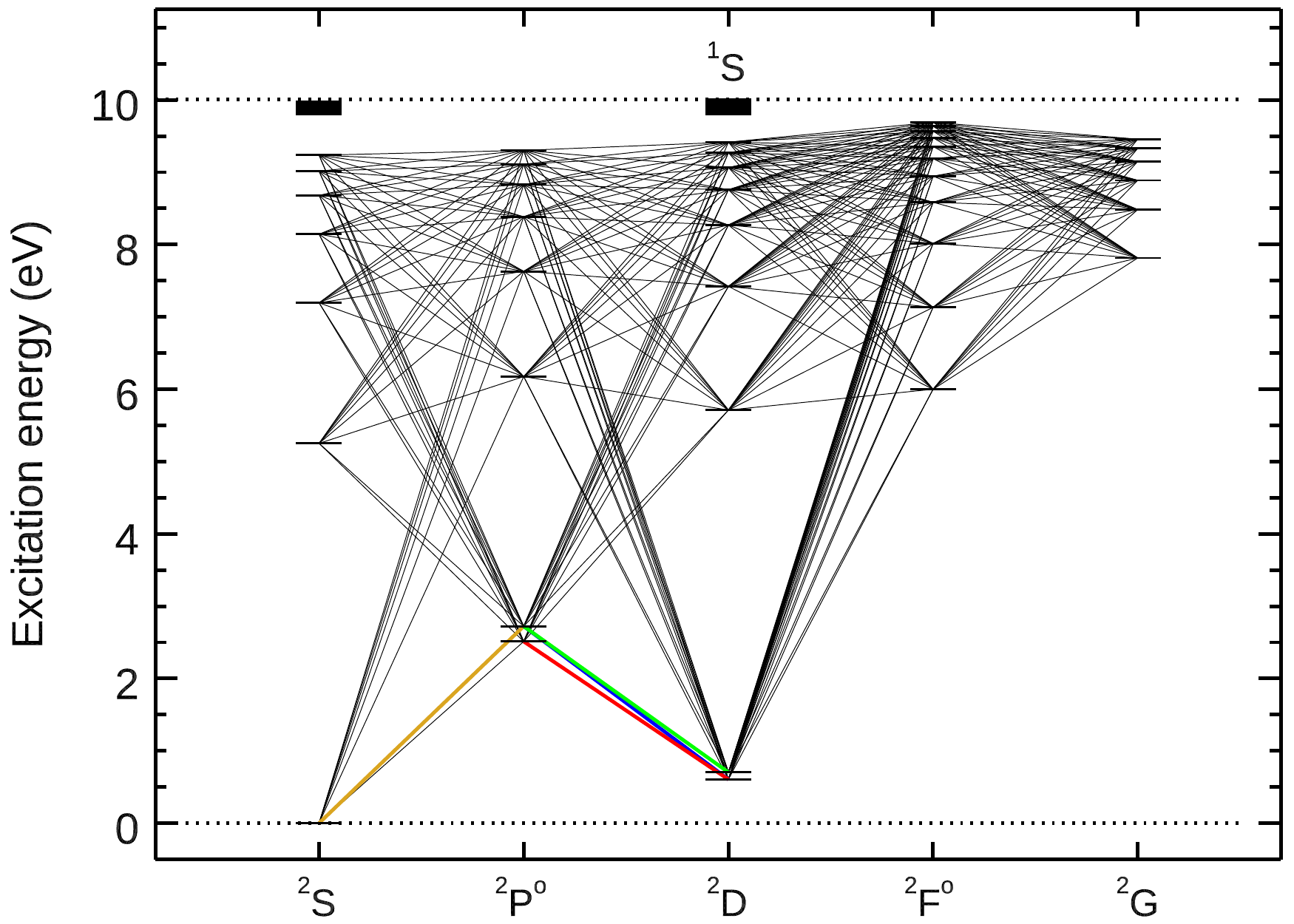}
\caption{The Grotrian diagram of \baii\ atom. Gold, green, blue, and red lines indicate the diagnostic \baii\ lines that we use in the solar abundance analysis at 4554, 5853, 6141, 6496\,\AA, respectively.}
\label{fig:ba_grot}
\end{figure}

The model atom of barium is constructed as follows. The energy levels for the \bai\ and \baii\ levels are extracted from the NIST database. Of these, we include eight energy states of \bai\ up to the energy of $2.86$\,eV, and all available levels of \baii\ up to $9.98$\,eV. Fine structure is retained for the three lowest terms of \baii: \Ba{6s}{2}{S}{}{} (ground state), \Ba{5d}{2}{D}{}{} ($\sim 0.65$\,eV), and \Ba{6p}{2}{P}{\circ}{} ($\sim 2.6$\,eV) (Table~\ref{tab:config}). Transitions between these terms are typically used in the barium abundance analysis of cool stars. Other levels are merged into terms, and their energy levels are represented by the weighted sum of the individual components (weighted by the statistical weights of the levels). In total, the model is comprised of $110$ states and is restricted by the ground state of \baiii\ at $15.2$\,eV for a total of $111$ levels. Note that no lines of \bai\ are observed in the spectra of FGK stars. For the Sun $n_\bai/n_\baii\approx10^{-4} - 10^{-6}$ (see Sect.~\ref{sec:nlte}), hence, \bai\ is a minority species. As such, a detailed treatment of the neutral stage is of no importance to the statistical equilibrium \baii, which is the majority species in these stars. Figure \ref{fig:ba_grot} depicts the energy levels in \baii\ and transitions among them in the form of a Grotrian diagram. We have colour-coded the four lines we use here as follows: gold represents the \baii\ 4554\,\AA\ line; green represents the 5853\,\AA\ line; blue represents the 6141\,\AA\ line; and red represents the 6496\,\AA\ line. As there are too many energy levels in the model atom to accurately depict without overlapping energy states, this figure should be used for qualitative assessments only.

The radiative bound-bound transitions for \baii\ were extracted from the \href{http://kurucz.harvard.edu/atoms/5601/gf5601.all}{Kurucz database}, 26.03.2017. We also compared the data with the \href{https://physics.nist.gov/PhysRefData/ASD/lines_form.html}{NIST database}. For the combined terms, the lines were merged and the transition probabilities co-added as described in \citet{Bergemann2012a}. The oscillator strengths of the four diagnostic lines used for the present analysis were extracted from the experimental transition probabilities presented in \citet{DeMunshi2015} and \citet{Dutta2016}. In total, the model contains $284$ spectral lines in the wavelength range from $1330$ to $202930$\,\AA. The transitions with $\log gf < -10$ are not included. Most of these lines are represented by nine frequency points, except for the four diagnostic lines that we use in the abundance analysis, i.e the lines at $4554$, $5853$, $6141$, and $6496$ \AA. These lines were represented with a profile containing $301$ frequency points. Several UV lines are rather strong. To test whether nine frequencies were enough to describe those strong lines, we ran a test using \multi. The departure coefficients from the model atom were compared with an atom that contained $100$ frequency points for two strong UV transitions at $2304.247$\,\AA\ and $2341.429$\,\AA. It was found that these transitions have no effect on the populations of the four barium lines of interest. This justifies the number of frequency points chosen for transitions that were not of interest to us for this study.
Damping by elastic collisions with hydrogen atoms are computed using the $\alpha$ and $\sigma$ parameters from \citet{Barklem2000} where available. When this information was missing, we used the Uns{\"o}ld approximation, which was scaled by $1.5$. The  wavelengths are taken from \citet{Karlsson1999}. We found that for three lines (NB: not the line at 4554\,\AA) there was a systematic offset to the solar spectrum. Once the solar spectrum was corrected for gravitational redshift they did match the observed line positions, although the 4554\,\AA\ resonance line has a slightly different shift. \citeauthor{Karlsson1999} underline that this line might be slightly shifted in their measurements, due to the isotopic mix they used and self-absorption in this strong line. This shift is however expected to be at most 1\,m\AA\ to the blue (Litz\'{e}n, private comm.) not sufficient to explain the remaining offset we observe. We discovered that the excess shift in this line was due to convective effects. We shifted the 4554\,\AA\ line by 2\,m\AA\ in 1D to the blue to match the observed position, whereas we did not need to shift the 3D profile.

We also introduce HFS and isotopic shifts. They were computed using the solar abundance ratios of the five barium isotopes, see \citet{Eugster1969}. The odd barium isotopes have non-zero nuclear spins that causes hyperfine splitting of the levels. The magnetic dipole and electric quadrupole constants for the five relevant energy levels were taken from \citet{Silverans1986}, and \citet{Villemoes1993}. The isotopic shifts are provided by \citet{VanHove1982} for the 5853 and 6141\,\AA\ lines, by \citet{Villemoes1993} for the 6496\,\AA\ line, and by \citet{Wendt1984} for the 4554\,\AA\ line. The diagnostic lines are hence represented by six to 15 HFS components. The complete HFS information for these lines can be found in tables in Appendix~\ref{apdx:hfs}. Oscillator strengths for these lines are computed from accurate experimental transition probabilities in \citet{DeMunshi2015} \citet{Dutta2016}.

The radiative bound-free data are computed using the standard hydrogenic approximation (Kramer's formula). This is appropriate since the first ionisation potential of \baii\ is at 10\,eV, and the energy levels, which may contribute to radiative over-ionisation at the solar flux maximum, have very low population numbers. Also on the basis of earlier studies with strontium \citep{Bergemann2012b}, which has a similar atomic structure, we do not expect that photo-ionisation is a significant non-LTE effect. In fact, earlier studies of barium in non-LTE showed that \baii\ a majority ion and is collision-dominated \citep[see][]{Mashonkina1999,Gehren2001,Bergemann2014}. In such ions, the statistical equilibrium is established by a competition of collisional thermalisation, photon losses in strong lines, and over-recombination. This will be discussed in detail in the next section.

One of the new features of our atom, compared to earlier studies mentioned above, is the treatment of collisions. In particular, we include the new quantum-mechanical rate coefficients for the inelastic collisions between the \baii\ ions and hydrogen atoms by \citet{Belyaev2017,Belyaev2018}. 
To the best of our knowledge, the first study of barium that employs these detailed quantum-mechanical data for collisions with hydrogen was recently published by \citet{Mashonkina2019} for the purposes of treating isotopes. The present paper is the first time these hydrogen-collision data have been employed for full non-LTE modelling.
The data are available for 686 processes, and represent collisional excitation and charge transfer reactions, i.e. ${\rm Ba + H} \leftrightarrow {\rm Ba}^{+} + {\rm H}^{-}$. The rate coefficients are typically large and may exceed $10^{-8}\,{\rm cm}^3{\rm s}^{-1}$ in the temperature regime relevant to modelling the solar atmosphere. 
Note that here the process ionisation refers to the ion we are interested in and a free electron, i.e. ionisation: $\baii + {\rm H} \to \baiii\ + {\rm H + e}$, but the process ion-pair formation reads: $\baii + {\rm H} \to \baiii\ + {\rm H}^{-}$, that is, \baii\ loses its outer electron and it is bound with H. Its inverse process, mutual neutralisation, is the process when \baiii\ gains an electron from H$^{-}$. The same is valid when \baii\ is replaced by \bai\, and \baiii\ is replaced by \baii. 
Charge transfer reactions do not lead to a free electron, which is the case that is usually modelled by a Drawin’s formula \citep{Drawin1968,Drawin1969}. Excitation and ionisation by collisions with free electrons are computed using the \citet{vanRegemorter1962} and \citet{Seaton1962} formulae. A study of the impact of different collisional rates was presented in \citet[][Sect. 3.2]{Andrievsky2009}. No differences between these classical recipes were found. An earlier version of this model atom was used in \citep{Eitner2019}. We have since updated the oscillator strengths.

\section{Barium line formation}
\label{sec:formation}

\subsection{non-LTE effects}
\label{sec:nlte}

In terms of departures from LTE, \baii\  is a collision-dominated ion \citep[see][and references therein]{Gehren2001}. The ionisation potential is too high for photo-ionisation to play a significant role in the statistical equilibrium (SE) in FGK type stars. On the other hand, the term structure of the ion, with several very strong radiative transitions, favours strong effects caused by line scattering. In particular, there is radiative pumping at the frequencies of optically-thin line wings, $\tau_{\rm wing} < 1$, as long as at the line centre $\tau_{\rm core} > 1$. This mechanism acts predominantly at larger depths and leads to over-population of the upper levels of the transitions, at the expense of the lower states. On the other hand, strong downward cascades occur higher up in the atmosphere, where the strong line cores become optically thin, $\tau_{\rm core} < 1$. This mechanism de-populates the upper levels via spontaneous de-excitations and this downward electron cascade causes over-population of the lower-lying energy states. As the statistical equilibrium of \baii\ has been extensively studied in the literature \citep{Mashonkina1996, Mashonkina1999, Short2006, Mashonkina2008}, hence in what follows we will only describe the main features of the non-LTE line formation and discuss the differences with the earlier studies.

\begin{figure}
\includegraphics[width=\linewidth]{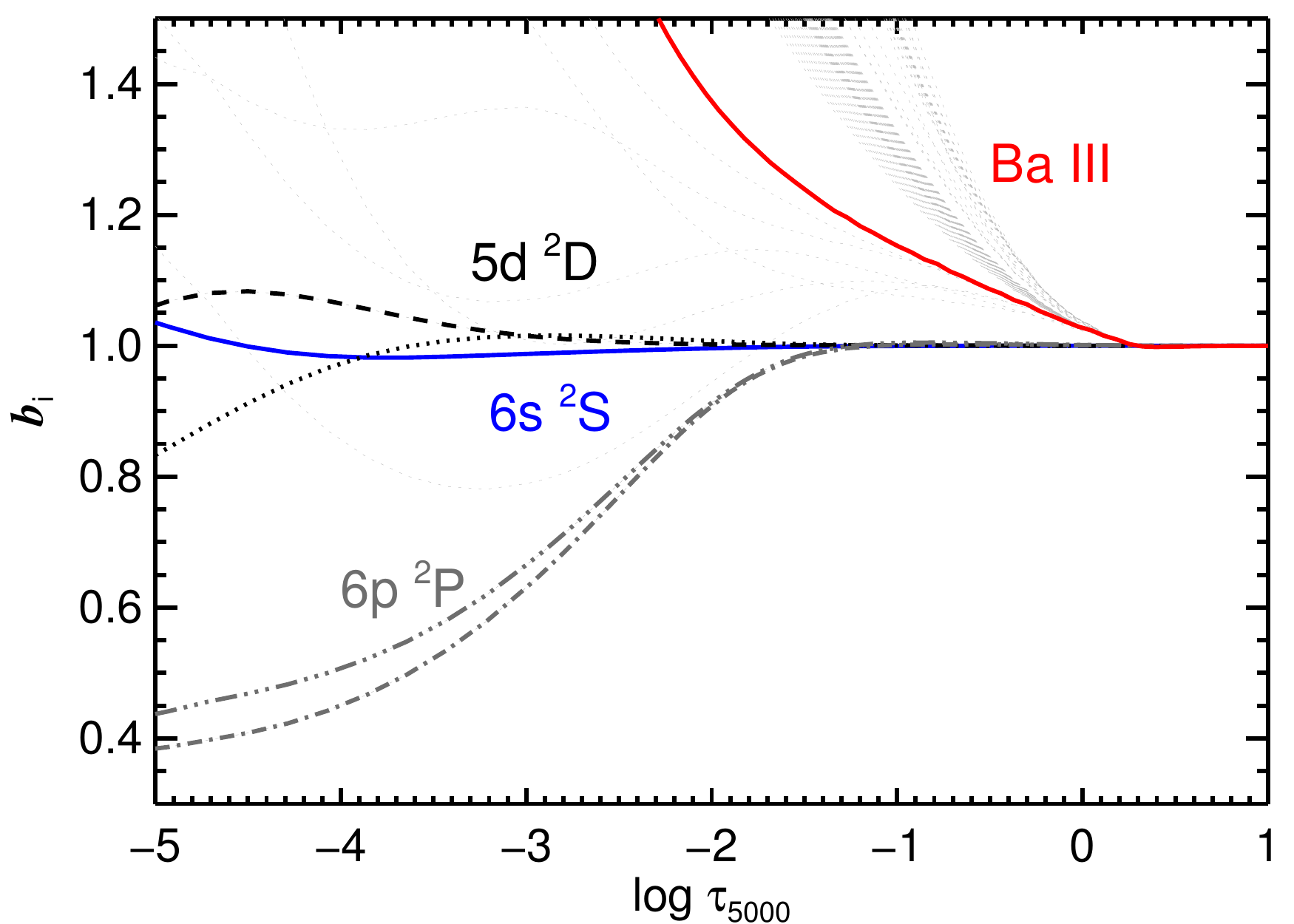}
\caption{Departure coefficients for the \baii\ levels as a function of the continuum optical depth $\opd$ computed using the reference barium model atom for the solar MARCS model atmosphere.}
\label{fig:deps}
\end{figure}


Quantitatively, this behaviour can be visualised as plots of the departure coefficients $b_i$\footnote{The departure coefficient is defined as the ratio of atomic number density for a given energy level $i$ computed in non-LTE to that of LTE, $b_i = \frac{n_{i,{\rm non-LTE}}}{n_{i,{\rm LTE}}}$.} as a function of the continuum optical depth at 5000\,\AA, $\opd$. Figure \ref{fig:deps} depicts the $b_i$ behaviour for the solar MARCS model atmosphere. To facilitate the comparison with the detailed study by \citet{Mashonkina1999}, we have chosen the same axis range as that paper. Thick coloured curves correspond to $5$ energy states, which are involved in the radiative transitions listed in Table~\ref{tab:config}: the ground state of \baii, \Ba{6s}{2}{S}{}{}, and the low-excitation terms \Ba{5d}{2}{D}{}{} and \Ba{6p}{2}{P}{\circ}{}. Thin grey dotted curves show all other energy levels of \baii\ in the model atom. It is interesting that despite major differences in the model and numerical methods, including the properties of the atomic model, model atmosphere, and the SE code, the agreement between our results and that of \citet[][see their Fig. 2]{Mashonkina1999} is very good. In particular, the \baii\ ground state is entirely thermalised throughout the full optical depth range and develops a very modest over-population only at $\opd < -4$. The first excited metastable state \Ba{5d}{2}{D}{}{} is also close to LTE due to strong collisional coupling with the \baii\ ground state, although minor departures in the atomic number densities, of the order of a few percent, are seen at $\opd < -3$ and higher up. The term \Ba{6p}{2}{P}{\circ}{} at $\sim 2.6$\,eV shows stronger deviations from LTE already in the deep layers, $\opd \sim -2$, where the non-LTE population of the level is only about $80\%$ of the LTE value (that is equivalent to $b_i = 0.8$). The pronounced depletion of the term is caused by photon losses in the lines, connecting the ground state and the lowest metastable state with \Ba{6p}{2}{P}{\circ}{}. This under-population increases outwards as the lines become optically thin. All energy levels above \Ba{6p}{2}{P}{\circ}{} are over-populated at $\opd < 0$, the effect that \citet{Mashonkina1999} attributed to radiative pumping. 

It is interesting to briefly discuss the importance of micro-physical processes in the SE of \baii. \citet{Andrievsky2009} suggest that photo-ionisation cross-sections is the main source of uncertainty in the SE of \baii. This is only true for \bai, however, no lines of the neutral atom are observed in the optical or infra-red spectra of FGKM stars \citep{Tandberg1964}. Over-ionisation of \bai\ has no effect on the population of \baii, as the ratios of number densities of two ionisation stages are $n_{\rm Ba I}/n_{\rm Ba II} \sim 10^{-4}$ at $\opd \approx 0$, and drops to $\sim 10^{-6}$ in the outermost atmospheric layers in the solar model. This ratio is even more extreme in the atmospheres of metal-poor stars. For example, for a model atmosphere of an RGB star with $\teff = 4600$\,K, $\logg = 1.6$, and $\feh = -2.5$, $n_{\rm Ba I}/n_{\rm Ba II} \sim 10^{-5}$ at $\opd \approx 0$, but approaches $\sim 10^{-10}$ close to the outer boundary at $\opd \approx -5$. On the other hand, photoionisation in \baii\  is not important. The ion has a very high ionisation potential, and its well-populated energy levels with low excitation potentials have ionisation thresholds in the far-UV, at $\lambda < 1000$ \AA, where radiative flux in  FGK-type stars is negligibly small.  
We also recomputed the departure coefficients assuming $\sigph/100$ and $\sigph\times100$. This is  a very conservative estimate of uncertainty in the cross-sections, when comparing to a very similar atom, Sr, for which detailed quantum-mechanical cross-sections are available from \citep{Bergemann2012a}. It was found that only the non-LTE populations of the \baiii\ ground state change, but none of the important \baii\ levels show any difference with respect to our reference model.

A more important ingredient for the SE of \baii\ seems to be the accuracy of the data for inelastic collisional processes, in particular, those between \baii\ and \hi\ atoms. \cite{Short2006} suggest that the non-LTE line profiles are invariant to a factors of $0.1-10$ changes in the rates of transitions caused by collisions (NB they used approximate analytical formulae to represent these data). This may hold for a limited range of stellar parameters. For example, in the case of the Sun, using the Drawin's recipe or QM data does not give significantly different results. However, it is known that metal-poor stars are sensitive to non-LTE effects \citep{Bergemann2014}. Therefore, it would be reasonable to assume that they would also be sensitive to different collisional recipes. This is particularly true in hydrodynamic model atmospheres as the decoupled non-local radiation field leads to a cooling in the outer regions of the temperature structure, relative to the equivalent 1D model \citep[see e.g.][their Fig.~1]{Gallagher2016}. We intend to explore this in the near future.

\subsection{Line formation in hydrostatic and inhomogeneous models}
\label{sec:lineformation}

\begin{figure*}[th]
\begin{center}
\includegraphics[width=\linewidth]{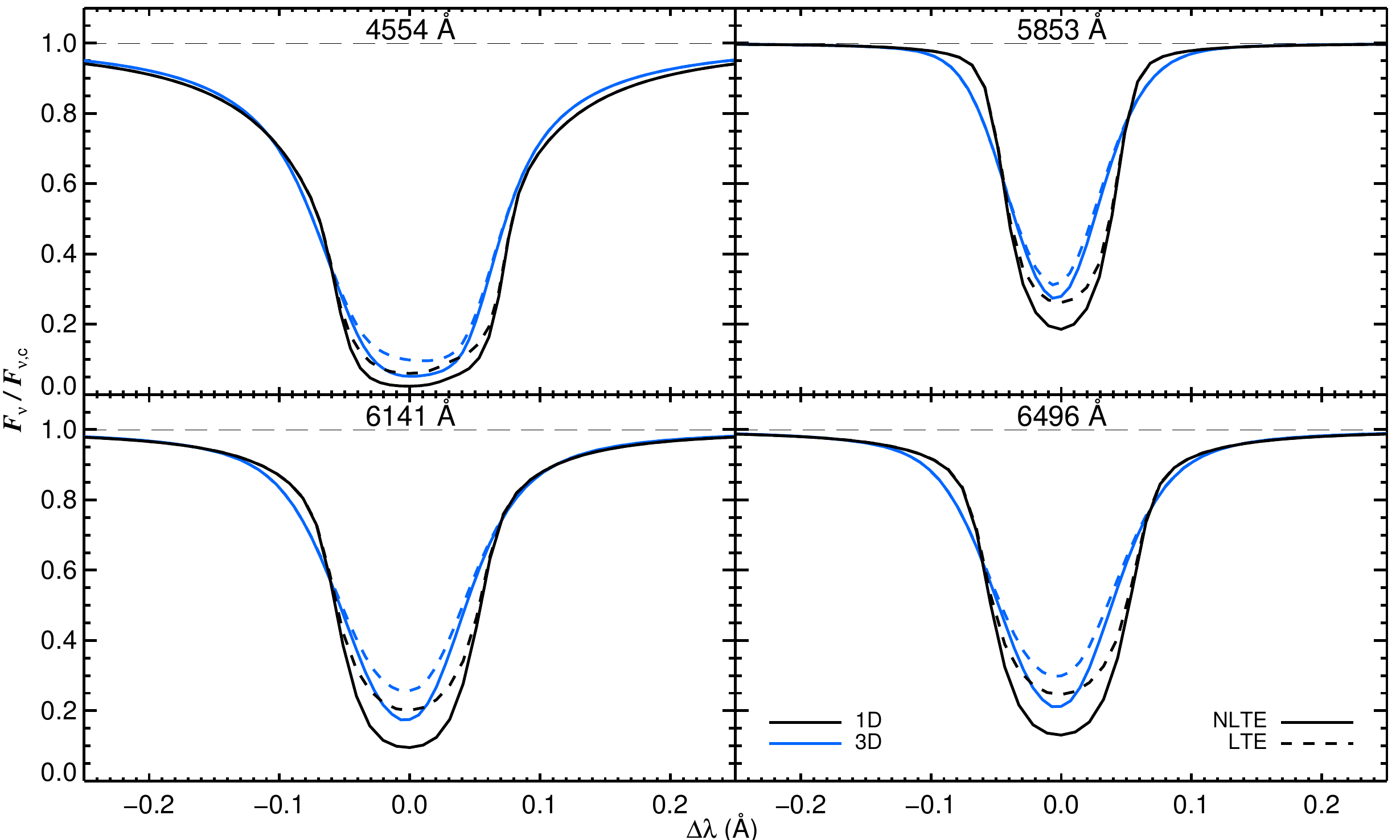}
\end{center}
\caption{1D (black) and 3D (blue) LTE (dashed-lines) and non-LTE (solid-lines) \baii\ line profiles at $\aba=2.26$\,dex. No extra broadening was added to any of the depicted lines. The 3D profiles are shown to be shifted relative to the 1D profiles. This is a natural consequence of the convective shifts indicative to a dynamic model atmosphere.}
\label{fig:LTENLTE}
\end{figure*}

We begin with the analysis of LTE and non-LTE formation of \baii\ lines in the 1D hydrostatic solar model.  We will then extend the analysis to radiative transfer with 3D inhomogeneous models.

As discussed in the previous section, the non-LTE effects in \baii\ are primarily dominated by line scattering. Consequently, deviations from LTE in the line source function are expected to be significant. Since the ratio of the departure coefficients for the lower $i$ and upper $j$ level of the transitions is below unity for the diagnostic \baii\ lines, $b_j/b_i < 1$, the ratio of source function to the Planck function \citep[see][for the derivation]{Bergemann2014}, also drops below unity. In other words, the source function in the line is sub-Planckian and the non-LTE lines profiles shall come out stronger than the LTE lines. In some cases \citep[e.g.][for Sr]{Bergemann2012a}, this effect is modulated by the change of the line opacity. However, since $\kappa_\nu \sim b_i$, and the population numbers for the lower levels of all \baii\ lines are essentially thermal throughout the line formation depths, the line opacity is very close to its LTE value. This simple analytical picture is confirmed by comparing the LTE and non-LTE line profiles (Fig. \ref{fig:LTENLTE}). The non-LTE effects are small and amount to the abundance difference of $-0.05$ (4554 \AA) to $-0.1$ dex (5853\AA). The other two \baii\ lines show similar behaviour.

Figure~\ref{fig:LTENLTE} demonstrates that the 3D profiles are asymmetric and also shifted blueward relative to the 1D profiles, which is expected. This is a natural result of the convective motions inside the 3D model, that the 1D model cannot replicate \citep{Lohner-Bottcher2018,Stief2019}. This is particularly obvious in the three subordinate lines, where the HFS has far less impact to the line shape than it does in the resonance line, where asymmetries are seen in both 1D and 3D. The 3D profiles, for a given barium abundance, are consistently weaker than their 1D counterparts, both in LTE and non-LTE. Therefore, positive abundance corrections are going to be needed to reproduce the same equivalent width, foreshadowing larger 3D LTE and non-LTE abundances over the 1D LTE counterpart. Like was shown in the hydrostatic case above, deviations from LTE should be significant because of line scattering. 

\begin{figure*}[!ht]
\begin{center}
\includegraphics[width=\linewidth]{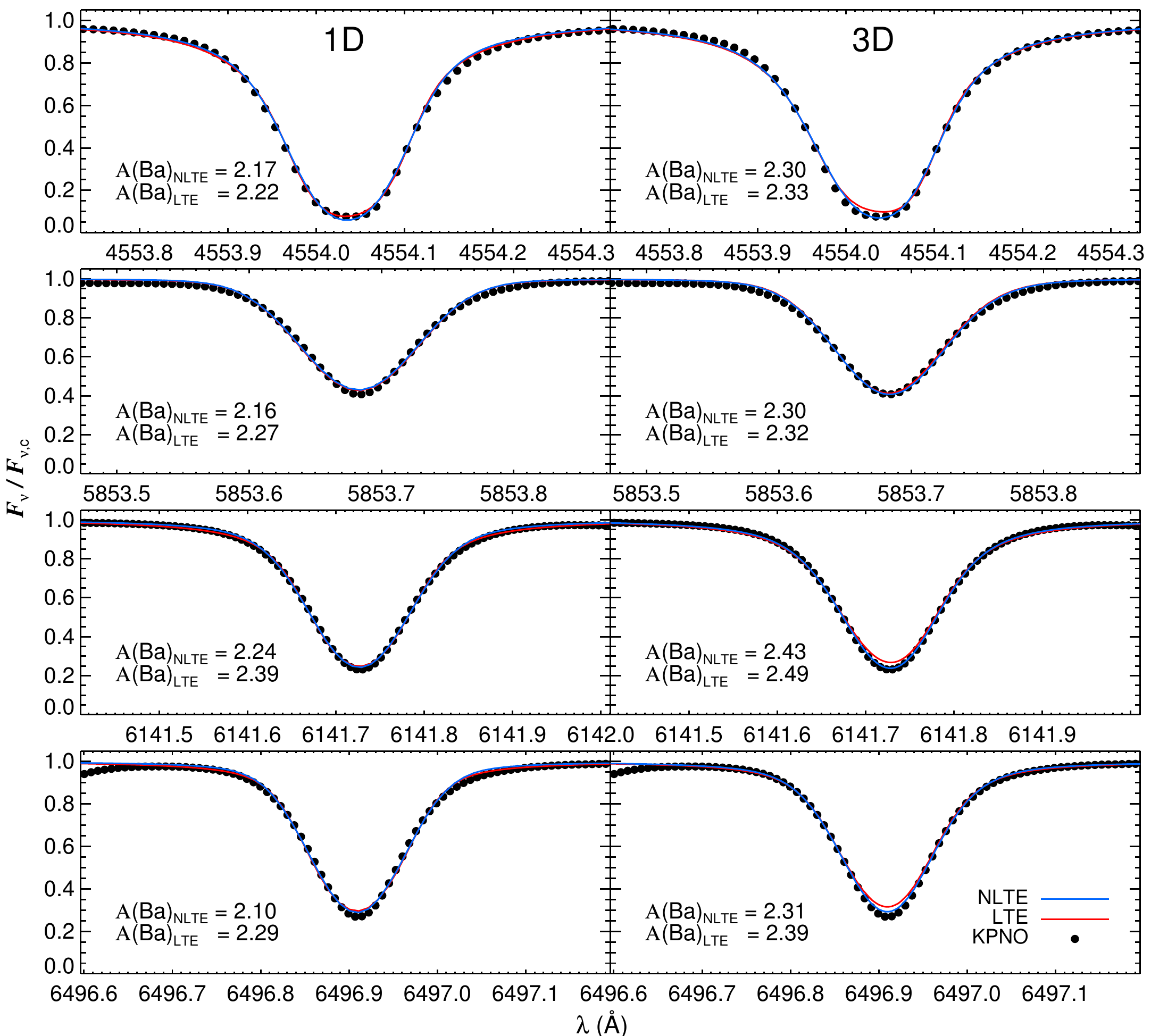}
\end{center}
\caption{Best fit \baii\ lines in 1D (left panel) and 3D (right panel) for both the LTE (red) and non-LTE (blue) cases. The 3D and 1D non-LTE profiles were computed using Table~\ref{tab:corrections}. Abundances  provided in the panels do not represent the final barium abundance, as line blends are excluded here. See Sect.~\ref{sec:results} and Table~\ref{tab:results} for details.}
\label{fig:bestfits}
\end{figure*}

The lines of \baii\ in the solar spectrum are strong, with EW's from 207\,m\AA\ (4554 line) to 68\,m\AA\ (5853 \AA) line.  As such, they are not only extremely affected by damping, but are also blended. The 4554 \AA\ line is blended by a \fei\ feature close the line core, but has little impact on the line. The subordinate lines also show some blending. \citet{Korotin2015} avoided, in particular, the 4554 \AA\ line in their analysis of FGK metal-poor stars in stars where $\feh>-1.0$. On the other hand, \citet{Grevesse2015} included the 4554, 5853, and 6496 \AA\ line in their analysis of the Sun. They also used the non-LTE corrections for \baii\ lines. We explore the effects of line blending and damping in Sect.~\ref{sec:results}.

The current photospheric solar abundance of barium derived by \citet{Grevesse2015} -- who applied non-LTE corrections to their 3D LTE abundance -- is $2.25\pm0.03\pm0.07$ (where errors represent the statistical and systematic errors, respectively). The abundance of barium in CI meteorites \citep{Lodders2003} is $2.19\pm0.03$. The main goal of this paper is to explore whether ab initio atomic data from physical experiments and detailed quantum-mechanical calculations are successfully able to describe the spectrum of the Sun. Accordingly, we now move on to report our 1D non-LTE, 3D LTE and 3D non-LTE corrections relative to the average barium 1D LTE abundance. 


\section{Computing abundance corrections}
\label{sec:method}

A grid of abundances for all four barium lines were computed using \multi. These were fit to the solar spectrum using a $\chi^2$ code that treats abundances, macroturbulent broadening and wavelength shifts as free parameters (see Sect.~\ref{sec:atom} for details of line shifts). The macroturbulences found ranged from $1.5-1.9\,\kms$. The code also normalises the fit to a local continuum for each line using two patches of spectrum either side of the line. We fixed the rotational broadening to $\vsini=1.6\,\kms$ \citep{Pavlenko2012}. The 1D LTE we attain represent the best statistical fit from this $\chi^2$ code. 

We computed the four barium lines in 1D non-LTE, 3D LTE and 3D non-LTE using three abundances; $\aba=2.17$, $2.27$ and $2.35$ - covering typical abundances reported in the literature for the solar 	barium abundance. The abundance corrections tabulated in Table~\ref{tab:corrections} were determined by fitting the grid of 1D LTE profiles to the 1D non-LTE, 3D LTE and non-LTE lines so that their equivalent widths matched. Unsurprisingly, it was found that the corrections from all three abundances were identical, hence corrections we provide are robust against the typical abundance range found by most studies on the solar barium abundance. 

\begin{table}[t]
\caption{1D non-LTE, 3D LTE, and 3D non-LTE (NLTE) abundance corrections, $\Delta$.}
  \begin{center}
  \begin{tabular}{l c c r}
    \hline
    Wavelength (\AA) & $\Delta_{\rm 1D\,NLTE}$ & $\Delta_{\rm 3D\,LTE}$ & $\Delta_{\rm 3D\,NLTE}$ \\
    \hline\hline
    $4554.033$ & $-0.05$ & $0.11$ & $0.08$ \\
    $5853.673$ & $-0.11$ & $0.05$ & $0.03$ \\
    $6141.711$ & $-0.15$ & $0.10$ & $0.04$ \\
    $6496.896$ & $-0.19$ & $0.10$ & $0.02$ \\
    \hline
  \end{tabular}
  \end{center}
  \tablefoot{Corrections are defined as $\Delta_D = A({\rm X})_D -A({\rm X})_{\rm 1D\,LTE}$, where $D$ is the 3D non-LTE, 3D LTE, or 1D non-LTE case.}
  \label{tab:corrections}
\end{table}


\section{Results}
\label{sec:results}

Our best fit profiles are compared with the solar flux profiles in Fig. \ref{fig:bestfits}. The best fit 3D lines are computed by \multitd\ using the corrections given in Table~\ref{tab:corrections}. The 3D non-LTE abundances are remarkably consistent, apart from the 6141\,\AA\ line, which is approximately $0.12$\,dex larger than the other three. The reasons for this will become apparent by the end of this section. Unlike in the 1D case, the best fit 3D LTE profiles show large deviations in the line cores, relative to their non-LTE counterparts, yet their equivalent widths remain very similar. 

The lines and abundances discussed this far have assumed that the barium lines are unblended in the solar spectrum. In reality this is not the case. In fact, the abundances all barium lines (particularly the $6141$\,\AA\ line) are dependent on line blending, as we now discuss.

\subsection{Line blending corrections}
\label{sec:blending}

The four barium lines we model here suffer from the effects of blending with other atomic and molecular species. Naturally, this impacts the abundances we derive when we assume that the line is clean. This is what was done when synthesising the lines with \multitd\ and \multi, as these codes do not currently possess the capability to synthesise blends. To examine this, we used the VALD3\footnote{\url{http://vald.astro.uu.se/}} database together with the barium line information extracted from the model atom to create new line lists for the 1D LTE spectrum synthesis code, MOOG\footnote{\url{https://www.as.utexas.edu/~chris/moog.html}}. This code was chosen as the interactive plotting tool makes recomputing synthetic spectra and fitting it with observed data very simple. Lines were computed with and without and the abundance of the clean barium line was adjusted until its line strength matched the blended line.

The \baii\ resonance line at $4554$\,\AA\ is the strongest line presented here. Naturally, it would dominate most of the line depression at this spectral region. It is found that when blends are included, the 1D LTE barium abundance must be reduced by $0.01$\,dex. The $5853$\,\AA\ line abundance was reduced by $0.03$\,dex. The $6141$\,\AA\ line was found to be severally affected by blending, as the abundance had to be reduced by $0.16$\,dex. The $6496$\,\AA\ barium abundance had to be reduced by $0.04$\,dex.

We will use these abundance corrections to determine the barium abundance in all four paradigms in Sect.~\ref{sec:baabundance}. First, however, it is important to determine how the barium lines are affected by systematic uncertainties, as we now present.

\subsection{Line damping uncertainties}
\label{sec:damping}

We have previously mentioned that differences in the radiative transfer solvers used by \multi\ and \multitd\ lead to extremely small systematic uncertainties in the barium abundance. These are small ($<0.01$\,dex) enough to be dwarfed by uncertainties associated with the van der Waals broadening parameters, now discussed. The barium lines synthesised here have varying line strengths. This means that they will react differently to uncertainties associated with the van der Waals line damping parameters, $\sigma$ and $\alpha$. If we conservatively assume that there is a $10\%$ uncertainty associated with the damping parameters tabulated in \citet{Barklem2000} then we can see how this affects the abundance measured in each line and derive a systematic uncertainty for each line. This does not affect the abundance corrections, as these systematics will affect all syntheses equally and so they cancel out, but it is important to determine any uncertainty associated with the abundances determined from these corrections.

We examined how both the cross sections ($\sigma$) and velocity exponent ($\alpha$) affect the line strength by varying both separately and computing new lines using \multi. It was found that varying $\alpha$ by $\pm10\%$ had no impact on the line strengths of any of the lines, and hence we cannot attribute any abundance uncertainty to this parameter. However, a $10\%$ uncertainty in $\sigma$ was found to affect the line strength in all four lines. The 4554\,\AA\ line is the strongest line measured (207\,m\AA). It is therefore reasonable to assume that this line will be the most affected by this damping parameter uncertainty. It was found that the barium abundance determined from this line varies by $\pm0.03$\,dex. The 5853\,\AA\ line is the weakest line in this study (68\,m\AA). As such the uncertainty we assign to the van der Waals parameter of $\pm10\%$ leads to a change in abundance of only $0.01$\,dex. The 6141\,\AA\ line is the second strongest line measured (126\,m\AA). This leads to abundance variations of $\pm0.03$\,dex. Finally, the 6496\,\AA\ line is also fairly strong in the solar spectrum (102\,m\AA). The uncertainty we assign the damping parameter varies the barium abundance found in this line by $\pm0.02$\,dex. The list of associated abundance uncertainties can be seen in  column four of Table~\ref{tab:results}.

\begin{table*}[th]
\caption{Abundances, associated error estimates and abundance corrections due to line blending.}
  \begin{center}
  \begin{tabular}{l c c c c c c c c c c r}
    \hline
    &&&&&&& \multicolumn{2}{c}{LTE} && \multicolumn{2}{c}{non-LTE} \\
    \cline{8-9} \cline{11-12}
    Wavelength (\AA) & $A({\rm Ba})_{\rm 1D,LTE}$ & $\Delta_{\rm blend}$ & $\sigma_{\rm BPO}$ & $\sigma_{\rm blends}$ & $\sigma_{f{\rm -value}}$ & $\sigma_{\rm total}$ & $A({\rm Ba})_{\rm 1D}$ & $A({\rm Ba})_{\rm 3D}$ &&  $A({\rm Ba})_{\rm 1D}$ & $A({\rm Ba})_{\rm 3D}$ \\
    \hline\hline
    $4554.033$ & $2.22$ & $-0.01$ & $\pm0.03$ & $\pm0.00$ & $\pm0.00$ & $\pm0.03$ & $2.21$ & $2.32$ && $2.16$ & $2.29$ \\
    $5853.673$ & $2.27$ & $-0.03$ & $\pm0.01$ & $\pm0.00$ & $\pm0.01$ & $\pm0.01$ & $2.24$ & $2.29$ && $2.13$ & $2.27$ \\
    $6141.711$ & $2.39$ & $-0.16$ & $\pm0.03$ & $\pm0.02$ & $\pm0.01$ & $\pm0.04$ & $2.23$ & $2.33$ && $2.08$ & $2.27$ \\
    $6496.896$ & $2.29$ & $-0.04$ & $\pm0.02$ & $\pm0.00$ & $\pm0.01$ & $\pm0.02$ & $2.25$ & $2.35$ && $2.06$ & $2.27$ \\
    \hline
  \end{tabular}
  \end{center}
  \tablefoot{Column 2 tabulates the 1D LTE abundances found when the lines were assumed to be clean. They are consistent with the profiles depicted in  Fig.~\ref{fig:bestfits}. Column 3 presents the abundance correction associated with line blending. The final four columns present actual abundances of each line when the appropriate corrections were added to Cols. 2 and 3 from Table~\ref{tab:corrections}. The weighted averages calculated in Sect.~\ref{sec:baabundance} are computed using these abundances and the systematic errors tabulated in Col. 7 from the errors in BPO theory \citep{Barklem2000}, line blend uncertainties, and oscillator strengths, f , which are presented in Cols. 4, 5, and 6, respectively.}
  \label{tab:results}
\end{table*}

Uncertainties in line blends are also of concern when computing abundances. No uncertainty information is given by VALD, so we again conservatively assume that the $\log{gf}$ values of these lines have a $10\%$ uncertainty. The abundances attained from the 4554\,\AA\ line with and without line blending were virtually identical. As previously mentioned, this is because the resonance line dominates line depression around this spectral region. Accordingly, uncertainties in blended lines of $\pm10\%$ do not affect the barium abundance. While the 5853\,\AA\ line is the weakest line analysed, it suffers the least from blending. As such, there is also no sensitivity in barium abundance found from varying the blended lines. The blends around the 6141\,\AA\ line have a large impact on the barium abundance. Naturally, uncertainties in $\log{gf}$ lead to an uncertainty of $\pm0.02$\,dex. Therefore, the inclusion of blend uncertainties increases systematic uncertainty of the 6141\,\AA\ from $0.03$\,dex to $0.04$\,dex. Finally, the blending uncertainties around the 6496\,\AA\ line were not found to influence the barium abundance. A break down of the associated abundance uncertainties can be seen in column five of Table~\ref{tab:results}.

\subsection{Oscillator strength uncertainties}
\label{sec:gfuncertainties}

The oscillator strengths ($f$-values) of the four diagnostic lines used in the present study are taken from \citet{DeMunshi2015} and \citet{Dutta2016}. They also provide unique errors associated to each transition probability, which we convert in to oscillator strength uncertainties. The transition probabilities are extremely accurate, so the resulting uncertainties are very small. When these are included in our calculations we find that the propagated abundance error associated with the 4554\,\AA\ line is $\pm0.00$\,dex. The weakest line in our sample (5853\,\AA) has a propagated abundance uncertainty of $\pm0.01$\,dex. The most blended line (6141\,\AA) is found to have a propagated abundance uncertainty of $\pm0.01$\,dex. Finally, the 6496\,\AA\ line has an associated abundance error of $\pm0.01$. The associated abundance uncertainties for each line can be found in column six of Table~\ref{tab:results}.

\subsection{Solar barium abundance}
\label{sec:baabundance}

We now present the barium abundance in four paradigms: 3D non-LTE, 3D LTE, 1D non-LTE and 1D LTE. We correct the 1D LTE abundances using the corrections in Table~\ref{tab:corrections} and Sect.~\ref{sec:blending}, and weight them by their uncertainties listed in Table~\ref{tab:results} using an inverse-variance weighted mean ($\sum{\omega_i\,X_i}$, where $\omega_i=\frac{1}{\sigma^2}$).

We report for the first time a full 3D non-LTE solar barium abundance of $2.27\pm0.02\pm0.01$, where errors given here are the systematic uncertainties and the random error determined as the standard deviation found in the line-to-line scatter of the 3D non-LTE abundances, which can also be seen in Table~\ref{tab:results}. This value is $0.08$\,dex larger than the meteoritic barium abundance of $2.19\pm0.03$ determined by \citet{Lodders2003}. We therefore find a photospheric abundance that is slightly larger than that reported from meteorites.

Using the same method just described, we find that $\aba=2.31\pm0.02\pm0.03$ in 3D LTE. Errors again represent the systematic and random errors, like above. When we use 1D model atmospheres and apply non-LTE physics to the post-process spectral synthesis we find that the 1D non-LTE barium abundance is $\aba=2.11\pm0.02\pm0.05$. Finally, when we derive the barium abundance using the classical 1D LTE approach we find that $\aba=2.24\pm0.02\pm0.02$. This abundance is similar to the 3D non-LTE abundance.

\section{Conclusions}
\label{sec:conclusions}

We have computed new values of the solar barium abundance based upon results computed using a new barium model atom that includes quantum mechanical inelastic collisional rate coefficients between \baii\ and hydrogen. We computed the barium abundance using a static 1D model atmosphere and provide 1D non-LTE, 3D LTE and 3D non-LTE corrections to this value in Table~\ref{tab:corrections}. Using these corrections we also present new solar photospheric barium abundances for the first time in full 3D non-LTE, as well as in 3D LTE, 1D non-LTE and 1D LTE. 

The summary of this work is as follows (NB that all abundances given below are done so by adding the corrections to each 1D LTE abundance and then calculating the inverse-variance weighted mean as described in Sect.~\ref{sec:baabundance}):

\begin{itemize}

\item The 3D non-LTE barium abundance was found to be $\aba=2.27\pm0.02\pm0.01$, which is $4\sigma$ larger the meteoritic abundance published by \citet{Lodders2003}. 
This may suggest uncertainties in the atomic data and/or that further physics is still missing from our analyses, such as the treatment of magnetic fields. On the other hand, Ba isotopic abundance anomalies are well-documented in CI meteorites \citep[e.g.][]{McCulloch1978,Arnould2007} and it is not clear whether the meteoritic value can be indeed directly compared to the solar photospheric estimate. Nevertheless, this value represents the best photospheric solar barium abundance available from the current state-of-the-art in spectral modelling. As a result, it provides a remarkably consistent abundance for the four diagnostic lines.

\item The 3D LTE abundance was determined as $\aba=2.31\pm0.02\pm0.03$. This value is larger than the meteoritic value given in \citeauthor{Lodders2003} and the 3D non-LTE abundance we determine, and the individual abundances are not as consistent.

\item The 1D non-LTE abundance of $\aba=2.11\pm0.02\pm0.05$ suggests that the barium abundance is depleted by $0.16$\,dex in the solar atmosphere relative to the full 3D non-LTE. The abundance is also smaller than the meteoritic result reported in \citeauthor{Lodders2003} and the inconstencies between lines are larger than they are in 3D non-LTE.



\end{itemize}

The 3D non-LTE and 1D LTE abundances are very similar for barium in the Sun, but larger than that given in \citet{Lodders2003}. 
Conversely, the 3D LTE abundance suggests that barium abundance is even larger in the Sun, while the 1D non-LTE abundance suggests barium is slightly lower than the meteoritic value. It is clear then that the inclusion or removal of realistic treatments of line formation physics or convection has opposing effects on the barium abundance; the former strengthens the barium lines and so depletes the barium abundance, while the latter weakens the barium lines and hence a larger barium abundance is required. Therefore, the exclusion of both physical processes in the 1D LTE paradigm masks each effect, providing similar values in each line as the actual values found in 3D non-LTE. Conversely, in our work on manganese we found that the 3D and Non-LTE effects do not cancel out, but rather the effects of Non-LTE are amplified in 3D calculations with hydrodynamical models.

The previous set of transition probabilities reported by \citet{Davidson1992} led to abundances values that were, in general, less consistent than those reported here and had larger uncertainties associated to them, with the 5853\,\AA\ line being most uncertain and most inconsistent with the other three diagnostic lines. The latest published transition probabilities in \citet{DeMunshi2015} and \citet{Dutta2016} represent the most accurate transition parameters published. As such, the barium abundances we find from each diagnostic line used here are all in very good agreement (once the blending corrections are included). 

We have presented this work as part of a larger series of papers designed to report on the development of the \multitd\ spectrum synthesis code. Up until now we have added new coding that allows it to read standard \stagger\ model atmospheres; include the effect of charge transfer between hydrogen and ions; compute flux data based on hard-coded quadrature schemes; and compute multi-component transitions caused by HFS or isotope splittings. Further physics and mathematical techniques will be added as the project progresses that will be presented in future papers in this paper series. We also plan to extend our work on barium within this paper series to include several metal-poor benchmark stars, where the present work will be important to the relative abundances we report.


\begin{acknowledgements}
This work made heavy use of the Max Planck Computing \& Data Facility (MPCDF) for the majority of the computations. This project was funded in part by Sonderforschungsbereich SFB 881 "The Milky Way System" (subproject A5) of the German Research Foundation (DFG) and by the Research Council of Norway through its Centres of Excellence scheme, project number 262622. SAY and AKB gratefully acknowledge support from the Ministry for Education and Science (Russian Federation), project Nos. 3.5042.2017/6.7, 3.1738.2017/4.6. BP is partially supported by the CNES, Centre National d’Etudes Spatiales.
\end{acknowledgements}

\bibliographystyle{aa}
\bibliography{../../../master}

\begin{appendix}

\section{HFS information}
\label{apdx:hfs}

This following tables tabulate the complete hyperfine structure information of the four barium diagnostic lines used in this study.

\begin{table}[h]
\caption{HFS informaion on the barium 4554\,\AA\ line.}
\begin{center}
\begin{tabular}{l c r}
\hline
$\lambda$\,(\AA) & Isotope & Relative strength \\
\hline\hline
4553.9980 & 137 & 0.1562 \\
4553.9985 & 137 & 0.1562 \\
4553.9985 & 137 & 0.0626 \\
4554.0010 & 135 & 0.1562 \\
4554.0015 & 135 & 0.1562 \\
4554.0020 & 135 & 0.0626 \\
4554.0317 & 134 & 1.0000 \\
4554.0317 & 136 & 1.0000 \\
4554.0332 & 138 & 1.0000 \\
4554.0474 & 135 & 0.4376 \\
4554.0498 & 137 & 0.4376 \\
4554.0503 & 135 & 0.1562 \\
4554.0513 & 135 & 0.0311 \\
4554.0537 & 137 & 0.1562 \\
4554.0542 & 137 & 0.0311 \\
\hline
\end{tabular}
\end{center}
\label{tab:hfs4554}
\end{table}

\begin{table}[h]
\caption{HFS information on the barium 5853\,\AA\ line.}
\begin{center}
\begin{tabular}{l c r}
\hline
$\lambda$\,(\AA) & Isotope & Relative strength \\
\hline\hline
5853.6831 & 137 & 0.0875 \\
5853.6851 & 135 & 0.0875 \\
5853.6865 & 137 & 0.1001 \\
5853.6865 & 137 & 0.0626 \\
5853.6875 & 135 & 0.1001 \\
5853.6875 & 135 & 0.0626 \\
5853.6875 & 137 & 0.3499 \\
5853.6875 & 137 & 0.0626 \\
5853.6880 & 137 & 0.0248 \\
5853.6880 & 138 & 1.0000 \\
5853.6885 & 135 & 0.3499 \\
5853.6890 & 136 & 1.0000 \\
5853.6890 & 135 & 0.0248 \\
5853.6890 & 135 & 0.0626 \\
5853.6895 & 137 & 0.1249 \\
5853.6899 & 135 & 0.1249 \\
5853.6899 & 134 & 1.0000 \\
5853.6904 & 137 & 0.1001 \\
5853.6914 & 135 & 0.1001 \\
5853.6934 & 135 & 0.0875 \\
5853.6934 & 137 & 0.0875 \\
\hline
\end{tabular}
\end{center}
\label{tab:HFS5853}
\end{table}

\begin{table}[h]
\caption{HFS information on the barium 6141\,\AA\ line.}
\begin{center}
\begin{tabular}{l c r}
\hline
$\lambda$\,(\AA) & Isotope & Relative strength \\
\hline\hline
6141.7183 & 137 & 0.0041 \\
6141.7202 & 137 & 0.0585 \\
6141.7222 & 137 & 0.0064 \\
6141.7231 & 135 & 0.0041 \\
6141.7231 & 137 & 0.3750 \\
6141.7231 & 137 & 0.0727 \\
6141.7246 & 135 & 0.0585 \\
6141.7251 & 137 & 0.2332 \\
6141.7251 & 137 & 0.0562 \\
6141.7266 & 137 & 0.1312 \\
6141.7266 & 137 & 0.0626 \\
6141.7261 & 135 & 0.0064 \\
6141.7271 & 138 & 1.0000 \\
6141.7271 & 135 & 0.3750 \\
6141.7271 & 135 & 0.0727 \\
6141.7280 & 136 & 1.0000 \\
6141.7290 & 135 & 0.2332 \\
6141.7290 & 135 & 0.0562 \\
6141.7295 & 134 & 1.0000 \\
6141.7300 & 135 & 0.1312 \\
6141.7305 & 135 & 0.0626 \\
\hline
\end{tabular}
\end{center}
\label{tab:HFS6141}
\end{table}

\begin{table}[h]
\caption{HFS information on the barium 6496\,\AA\ line.}
\begin{center}
\begin{tabular}{l c r}
\hline
$\lambda$\,(\AA) & Isotope & Relative strength \\
\hline\hline
6496.8979 & 137 & 0.0311 \\
6496.8989 & 135 & 0.0311 \\
6496.9014 & 137 & 0.1562 \\
6496.9019 & 135 & 0.1562 \\
6496.9062 & 137 & 0.4376 \\
6496.9067 & 135 & 0.4376 \\
6496.9102 & 134 & 1.0000 \\
6496.9102 & 136 & 1.0000 \\
6496.9102 & 138 & 1.0000 \\
6496.9160 & 135 & 0.0626 \\
6496.9165 & 137 & 0.0626 \\
6496.9175 & 135 & 0.1562 \\
6496.9185 & 137 & 0.1562 \\
6496.9204 & 135 & 0.1562 \\
6496.9219 & 137 & 0.1562 \\
\hline
\end{tabular}
\end{center}
\label{tab:HFS6496}
\end{table}

\end{appendix}

\end{document}